# Self-induced nonreciprocity from asymmetric photonic topological insulators

Sema Guvenc Kilic[1,#], Ufuk Kilic[1], Mathias Schubert[1,2], Eva Schubert[1] and Christos Argyropoulos[3,*]

[1]Department of Electrical and Computer Engineering, University of Nebraska-Lincoln, Lincoln, NE 68588, USA

[2] Solid State Physics and NanoLund, Lund University, P.O. Box 118, 22100, Lund, Sweden

[3] Department of Electrical Engineering, The Pennsylvania State University, University Park, PA 16803, USA

[#]sguvenckilic2@huskers.unl.edu and *cfa5361@psu.edu

## Abstract

Photonic topological insulators and self-induced nonreciprocity based on nonlinear effects in asymmetric structures have garnered increased attention due to their potential applications in quantum information technologies and advanced photonic integrated circuits. In this study, we combine these two fields and present asymmetric photonic crystal designs based on all-dielectric checkerboard structures forming topological interfaces. The resulting topological photonic structures lead to the emergence of a narrowband leaky mode in their linear transmission response accompanied by an edge state with topological protection. The incorporation of the nonlinear optical Kerr effect in the system results to spectral tunability and transmission nonreciprocity in the induced topological edge states as the input beam intensity is increased due to the introduction of asymmetric defects along the interface. We envision that the findings presented in our work will lead to tunable photonic topological insulators with nonreciprocal response that promise to provide new avenues for the development of next-generation photonic devices and quantum optical technologies.

# Introduction

Photonic crystals (PhCs) have captivated researchers with their ability to control the propagation of electromagnetic waves through optical structures, mirroring the behavior of electrons navigating the periodic potentials in semiconductor crystals[1]-[2]. Over the last few decades, various PhCs have been developed due to the advent of nanofabrication and utilized in myriad different applications, including optical switches[3]-[4], optical logic gates[5], and integrated optical circuits[6]-[7]. However, most PhCs suffer from dissipation and scattering losses mainly due to fabrication imperfections and material losses. Moreover, other general problems, such as achieving high field confinement and tunability in the electromagnetic wave propagation, cause further restrictions to their practical applications.

Recently, the emerging field of topological photonics was brought into prominence as a potential remedy to overcome these limitations. Generally, the interior region of a topological insulator solid-state material behaves like an electrical insulator, while its surface can sustain conducting states, meaning that the electrons can only move throughout the material's surface. In photonics, Haldane and Raghu translated the theoretical findings of the topological insulator materials' unique abilities to new optical designs based on photonic crystals[8]-[9] which were later experimentally demonstrated[10]. Specifically, the incorporation of two PhCs in a composite system leads to the creation of the so-called photonic topological insulators (PTIs). The realized interface between the PhCs offers an efficient mechanism to harness their topological properties, yielding the development of new waveguiding systems with high signal-to-noise ratios due to decreased scattering loss[11]. Thus, PTIs have emerged as promising optical platforms, offering unprecedented opportunities for applications in quantum information technologies, waveguide systems, and the next generation of photonic integrated circuits, to name a few[12]-[15].

Hence, designing a PTI system from two different PhCs that share a common energy band diagram with an overlapping photonic band gap provides a solid route to control the light propagation. The topological properties of both PhCs are different and determined by the topological invariance metric that plays a fundamental role in PTI design. Depending upon the number of dimensions involved in the PhC, the topological properties can be determined using various methods including Chern number and Zak Phase calculations. For most two-dimensional (2D)[15]-[16] and all three-dimensional (3D)[17]-[19] PhCs, the appropriate topological invariant metric is the Chern number[20] that provides information about the number of excited topological edge states. The topological properties of one-dimensional (1D)[21]-[22] and several alternative 2D[23]-[25] photonic crystal systems are usually determined by the so-called Zak phase[26]. Numerous theoretical and experimental studies demonstrated that topological edge states can be observed regardless of impurities[27]-[28], imperfections[29], defects[30]-[33], and disorders[34]-[36] in various photonic crystal systems.

However, unidirectional electromagnetic wave propagation, which is directly related to breaking the Lorentz reciprocity theorem, is challenging to be achieved with PTI systems. This property will be useful for various applications, such as nonreciprocal transmission filters[37]-[38], optical diodes[39], circulators[40], isolators[41]-[42], and optical resonators[43]-[44]. While reciprocal 'spin locked' unidirectional edge modes propagating along the topological interface have been demonstrated when illuminating the edge state out of plane by using circular polarization, nonreciprocal transmission response, especially for in-plane linear polarized illumination, is more challenging to be realized. To tackle this problem, we need to engineer the PTI systems with the goal of generating a nonreciprocal response. Note that one of the proposed techniques to achieve nonreciprocity utilizes the Faraday effect in magneto-optical media that always requires an external magnetic field bias[45]-[48]. This method creates an asymmetric change in the permittivity tensor of the magneto-optical material by applying a strong magnetic field that requires bulky magnets and usually leads to high energy dissipation[49]. These disadvantages make such nonreciprocal systems impractical to operate at optical frequencies.

In recent years, the material's inherent nonlinear optical properties have been investigated as alternative approach to achieve strong nonreciprocal response in photonic systems[50]-[57]. By utilizing the nonlinear properties of optical materials, self-induced nonreciprocity has gained increased attention since these systems do not require external bias and work for pulsed illumination. In an analogous way, as the permittivity tensor is modified by applying a strong magnetic field in the case of magneto-optical materials, the materials' nonlinear permittivity part depends on the local field intensity. In the currently investigated topological photonic systems with asymmetric geometries, an increase in the intensity of the incoming light contributes to a shift on the edge mode frequency due to enhanced third-order Kerr nonlinearity. The existence of asymmetric resonant frequency shift, when forward and backward illuminations are considered, is counted as evidence of self-induced nonreciprocity[58],[59].

More specifically, we introduce a 2D photonic topological insulator system utilizing simplified photonic crystal designs featuring all-dielectric checkerboard structures. Through the introduction of a simple 2D lattice mismatch by performing a half-lattice constant shift along both 2D directions of the unit cell, we engineer a superlattice structure realizing a topological interface. The presented design leads to the emergence of a narrow leaky wave mode within the photonic topological gap that appears in the transmission spectrum. Both PhCs that create the interface possess identical dispersion diagrams. However, they have different Zak phases which are calculated by using the electric field distributions at high symmetry k-points yielding the band inversion. Therefore, the proposed composite design leads to a robust topological edge state that can be spectrally controlled and tuned due to the material's inherent Kerr nonlinearity. Moreover, we demonstrate that increasing the incident beam intensity results in a red shift in the leaky edge mode transmission resonance response. We also investigate the self-induced nonreciprocity due to the introduction of defects along the interface of the photonic topological system. The aforementioned defects can be air holes or dielectric pillars leading to geometric asymmetry. When the Kerr nonlinearity of the material is considered, strong and tunable self-induced nonreciprocal light transmission is realized at the induced edge mode along the interface.

Finally, it is worth noting that nonreciprocal transmission via nonlinear effects (e.g., Kerr nonlinearity) has been theoretically explored before in topological systems based on ideal lossless coupled ring lattices that induce topological phase transitions manifested as 'self-induced' topological solitons[41]. However, the physics and results in this previous work are substantially different compared to the currently presented practical (all losses are included in our configurations), integrated (planar geometry), and simple to fabricate (i.e., more practical) photonic systems based on tunable edge modes excited at nonlinear checkerboard photonic crystal interfaces.

Hence, we demonstrate self-induced nonreciprocal edge states formed along topological photonic interfaces. Such functionality is important, since only at the interface is it possible to realize the topologically protected and robust against disorder edge modes that are among the most distinguished properties of topological photonic platforms. To be able to extend the concept of self-induced nonreciprocity through nonlinearity to such topologically protected edge modes reveals new physics relevant to the emerging field of nonlinear topological photonic systems[60] that can lead to significant performance advantages in integrated photonics and new tunable nonreciprocal functionality with topological protection.

## Results and Discussion

**Topological Interface**

In this study, we realize a topological interface based on checkerboard photonic crystals to investigate nonreciprocity and tunable edge state propagation. The step-by-step details on how the topological interface is formed are presented in **Figure 1**. The photonic crystal checkerboard pattern consists of air and dielectric sections. Each face of the periodic pattern has an equal length with a lattice constant of *a*. Additionally, it is important to note that the permittivity of the dielectric material is taken equal to 11.56, similar to silicon (Si) at near-IR frequencies[61]. While Si losses at near-IR are negligible, the effect of Si losses in the visible and other realistic materials are considered in the Supplementary Material Sections 1-3. In order to create a photonic interface, the sequential translation operations are applied towards both x- and y-directions by shifting them along half lattice constant (*a*/2) (see **Figure 1**(a)-(c)). Hence, this two-step geometrical translation operation creates a lattice mismatch, resulting in a distinct interface between two PhCs (see **Figure 1**(d)). One side of the interface, referred to as Photonic Crystal-1 (PhC-1), consists of a repetition of air blocks surrounded by the dielectric material blocks. On the other side, Photonic Crystal-2 (PhC-2) is characterized by the repetition of dielectric blocks surrounded by air blocks. Such interface design is expected to achieve localization of the incident wave along its length, leading to leaking resonant transmission modes in the linear optical response[25]. Moreover, some of these modes can be topologically protected which is extremely important property for their use in several applications including quantum information systems, photonic integrated circuit devices, next-generation sensing, and waveguide systems[12]-[14],[62]-[65], where propagation of energy robust to backscattering from defects and other imperfections is required.

While the fundamental study of such resonant modes plays a crucial role in the experimental realization of the above-mentioned applications, boosting their signal-to-noise ratio and controlling the spectrum of these resonances are additional challenges that are seldomly discussed in literature. Even more importantly, we demonstrate the integration of defects at the interface of the proposed topological photonic system (see **Figure 1**(e)) which will lead to self-induced nonreciprocal response due to the geometric asymmetry and inherent nonlinear properties of the presented design.

**Linear Optical Response**

By using full-wave electromagnetic simulations based on COMSOL Multiphysics, the linear transmission and reflection characteristics of each individual photonic crystal and both forming the interface are investigated. To extract these responses, we employ a multiple port network, where the scattering parameters (S-parameters) are defined and computed[66] when multiple unit cells (24 in this case) of the proposed checkerboard structure are employed. While the propagation vector is aligned along the x-direction, the wave is linearly polarized with an electric field towards the out-of-plane z-direction. In **Figure 2**(a) and (b), the transmission and reflection spectra of both PhC-1 and PhC-2 designs are presented, respectively, where the lattice constant is chosen to be $a$=137 nm. We observe a clear photonic band gap region where the reflectance is 100% within a broad bandwidth of 130 THz. It is important to note that the photonic band gap region is independent of the unit cell number utilized in the design of PhC-1 and PhC-2. If the number of unit cells is increased, more Fabry-Perot oscillations will be observed outside of the band gap without affecting its bandwidth. Hence, the proposed photonic crystal design operates similar to an optical bandpass filter with response that can be tuned as a function of the lattice constant parameter $a$. More interestingly, when an interface is realized, we observe the emergence of a narrowband leaky transmission mode within the photonic band gap, as depicted in **Figure 2**(c) and its inset in **Figure 2**(d).

The resulting photonic topological insulator structure induces a high-quality factor (Q-factor) resonant leaky transmission mode in the linear optical response. The Q-factor of the photonic system is calculated using the ratio of the resonance frequency to its bandwidth. As seen from **Figure 2**(d), the resonance frequency and the bandwidth are approximately $422.4\ THz$ and $1\ GHz$, respectively. The ratio of these two values gives the Q-factor of our photonic design. This photonic system has a large Q-factor value approximately $4 \times 10^5$ mainly due to minimal energy and scattering losses. The Q-factor will drop when realistic losses are considered but still remain high, as it is shown in the Supplementary Material. To demonstrate that the presented response can be tuned to different frequencies, we extract the effect of the lattice parameter $a$ on both the full-width half maxima (FWHM) ($f_{FWHM}$) and central ($f_C$) frequencies of the realized photonic band gap (see **Figure 2**(e)) of each photonic crystal. Based on extensive simulations, we derive fitting relations for both $f_{FWHM}$ and $f_C$ as functions of the lattice parameter $a$ given as follows: $f_{FWHM} = 1.79 \times 10^4 a^{-1}\ THz.nm$ and $f_C = 5.9498 \times 10^4 a^{-1}\ THz.nm$, respectively, where the star in **Figure 2**(e) represents the lattice parameter used to derive the remaining results in the same figure. Interestingly, the presented photonic topological system is scalable, i.e., it can operate in various frequencies based on the unit cell lattice parameter $a$ value.

**Topological Properties: Zak Phase Calculations**

The obtained transmissive narrowband leaky mode generated in the photonic band gap region is an important advantage of the current design compared to relevant previous works, where the induced topological edge modes are dissipated by Ohmic losses due to the use of metals[67]-[68]. In order to gain a deeper understanding of the physics behind this leaky mode and investigate the corresponding energy band diagram and eigenmode characteristics, we employ simulations using the eigenfrequency solver in COMSOL Multiphysics. Given the very similar geometries of PhC-1 and PhC-2 unit cells, we expect that their corresponding dispersion diagrams will be identical. **Figure 3**(a) and (b) show the reduced 1D dispersion diagrams at fixed $k_y = 0.25\pi/a$ for both PhC-1 and PhC-2 unit cells, respectively. Note that frequency and k-vector are normalized to the lattice constant $a$ in all dispersion diagram plots. On one hand, the proposed PhCs indeed share the same dispersion diagram. On the other hand, due to the distinct inversion centers defining the origins of PhC-1 and PhC-2 unit cells, these two photonic crystals can potentially manifest different topological features. Thereby, we calculate the Zak phase values for both PhCs which serve as invariants that characterize the topological band properties. It basically represents the Berry phase acquired by a particle during its adiabatic motion across the Brillouin zone center and its boundaries[69]. Typically, for a photonic crystal system with inversion symmetry, the Zak phase of an isolated band that belongs to the unit cell dispersion diagrams is quantized as either 0 or $\pi$. This quantization is tied to the spatial-field symmetries of the Bloch waves[69]. In addition, the complete dispersion diagram ($k_x$ and $k_y$ vectors vary) of both PhCs with their shared photonic band gap is demonstrated in **Figure 3**(c), where the schematic representation of the first Brillouin zone is depicted as inset.

To determine the Zak phase of the first band (blue solid lines in **Figure 3**(a) and (b)) in the 1D reduced dispersion diagrams for PhC-1 and PhC-2, we took into account two high symmetry k points at A ($k_x = -\pi/a$) and B ($k_x = 0$). These selected points are indicated with magenta solid circles over the first band (blue solid line) depicted in **Figure 3**(a) and (b). The 2D cross-section color density plot of the induced electric field distribution for both photonic crystals at points A and B are shown together with the corresponding unit cells as insets in **Figure 3**(a) and (b), respectively. Upon comparing the $|\widetilde{E_z}|$ electric field values between PhC-1 and PhC-2 at points A and B, we observe significant differences at point A, whereas they are similar at point B. Hence, the corresponding quantized Zak phase values are computed for PhC-1 and PhC-2 using the following approach:

- if $|\widetilde{E_z}|$ is zero for one of the high symmetry points (either point A or B) at $x = 0$ and the other symmetry point is nonzero, the Zak phase is $\pi$.

- if $|\widetilde{E_z}|$ is nonzero for both high symmetry points (A and B) at $x = 0$, the Zak phase is 0. The electric field at the high symmetry points (for $k_x = -\pi/a$ and $k_x = 0$) is calculated by using the following formula[70]-[71]:

$$\widetilde{E_z}(x = 0) = \int_0^a \frac{1}{a}\widetilde{E_z}(x = 0, y)e^{-ik_y \cdot y} \, dy. \tag{1}$$

Numerical calculations for obtaining the equivalent $\widetilde{E_z}$ value over the entire unit cell area at points A and B, selected at the first band of each unit cell dispersion diagram, are the main steps before determining the Zak phase values of both PhCs. Our simulations reveal that while their unit cell dispersion diagrams are identical, their corresponding Zak phases are different. This intriguing finding suggests the possibility of edge states emerging at the interface between PhC-1 and PhC-2.

**Edge State Formation at the Interface**

The complete dispersion diagram of the proposed photonic crystal composite structure with the topological interface is obtained from eigenfrequency simulations and presented in **Figure 4**(a). While the grey-colored sections of the dispersion diagram indicate the bulk region, the solid magenta line corresponds to the dispersion band of the induced edge state. To theoretically demonstrate the topological waveguide characteristics of the proposed photonic topological interface, a photonic crystal structure is designed composed of 24 unit-cells along the x-direction and 12 unit-cells along the y-direction. Within this framework, two different interface shapes are studied: (i) straight line and (ii) step-shaped with two sharp corners, which are schematically illustrated on the left sides of **Figure 4**(b) and **Figure 4**(c), respectively.

Since the presented photonic topological insulator platform is designed by using dielectric materials with a real dielectric permittivity of 11.56 (emulating silicon at near-IR wavelengths), the light propagation along the interface does not suffer from ohmic losses. Note that the detrimental losses of metallic (plasmonic) materials hinder them from being utilized in such topological photonic systems, although plasmonic materials can strongly localize and enhance the incident beam at nanoscale regions[68]. In this study, maintaining sufficient contrast between the refractive index of air and the chosen dielectric material combined with the topological edge state formation result in the mitigation of scattering in the light propagation along the interface (see **Figure 4**(b) and **Figure 4**(c)). Light coupling to the aforementioned two interfaces is successfully demonstrated in **Figure 4**(b) and **Figure 4**(c) inlaid with their schematic representations. We also thoroughly calculate and present the electric field distribution induced at the edge state of the proposed topological photonic structure in Section 4 of the Supplementary Material.

The investigation of unidirectional light coupling at the interface when excited by circular polarized waves (spin locking) is another important feature of the theoretically predicted topological edge state. Thereby, we introduce right (left) handed circularly polarized (R(L)CP) dipoles at the topological interface to mimic the out-of-plane excitation of circularly polarized light sources. **Figure 4**(d) illustrates the propagation of light through the interface between two PhCs, accompanied by their schematic representations. In these simulation scenarios, right-handed circularly polarized (RCP) or left-handed circularly polarized (LCP) light is excited at the center of the interface between PhC-1 and PhC-2. Hence, a point dipole with an amplitude of $\vec{P} = (1, \pm i, 0)$ is placed along the interface, where the plus (minus) sign represents LCP (RCP). We observe that, at the determined edge mode frequency, the light propagates along $\pm x$ direction across the interface depending on the polarization handedness, as depicted in **Figure 4**(d). Moreover, the excitation of circularly polarized light (operating at the edge mode frequency) is strongly confined along the interface between PhC-1 and PhC-2, as depicted by the normalized electric field distributions. The chirality of incident electromagnetic radiation coupled along the interface demonstrates the preferential direction of light propagation at the topological edge state. The presented simulations align well with the edge dispersion characteristics derived from eigenfrequency modeling and confirm the spin locking of the realized topological edge mode.

Next, in order to investigate the spectral tunability of the realized topological edge state of the presented composite photonic structure, the frequency of the induced edge mode is computed for different lattice constants ($a$) and plotted in **Figure 5**(a). Hence, we demonstrate that while the edge state characteristics remain prevalent, the lattice constant can be utilized to spectrally tune the frequency of the induced topological edge mode. The correlation between the lattice constant and the corresponding frequency of edge mode maxima (**Figure 5**(b)) is found as $f_{max} = 1.0208 \times 10^5 \, a^{-1} \, THz.nm$. The details of incident beam characteristics employed in the dispersion diagram computations are presented in the schematic illustration of the photonic interface in **Figure 5**(b).

Note that the topological edge mode will exist even if the photonic crystals have defects along their interface, as long as their topological invariance does not change. Interestingly, the defect presence can be utilized to both spectrally tailor and control the edge mode characteristics. To demonstrate this unique ability, we introduce two different defect geometries at the interface of our photonic system: (i) dielectric pillars at the center of the checkerboard air section and (ii) air holes at the center of the checkerboard dielectric section (see the inset schematic of **Figure 6**(a)). **Figure 6**(a) shows the topological edge mode dispersion band diagrams of the proposed photonic structure with dielectric defects/no air defects (blue line), air defects/no dielectric defects (green line), and without defects (red line). It can be clearly seen in **Figure 6**(a) that the defects presence in the photonic crystal system detunes the edge mode frequency but, interestingly, does not diminish the edge mode existence. It should be stressed that both dielectric and air defects have radius of 45 nm, a value that can adequately fit inside the currently used photonic crystal unit cell, leading to substantial geometrical asymmetry. **Figure 6**(b) demonstrates the normalized electric field confinement at the interface in the presence of only air (green) or dielectric (blue) defects, and without defects (red) in the photonic crystal system. We observe that strong electric field confinement at the interface is still prevalent in the presence of defect formation (see magenta dashed line in **Figure 6**(b) that highlights the interface), as expected due to the topological protection of the induced edge mode.

**Self-induced nonreciprocity due to nonlinear Kerr effect**

Previously, nonlinear effects were utilized for the activation of self-induced nonreciprocity in both plasmonic nanoantennas and micro-ring resonators[58],[72]-[73], acoustic and elastic metamaterial designs[74], and epsilon-near-zero metamaterials[75]. In most cases the nonlinear Kerr effect was utilized, which provides an ultrafast tuning mechanism. Briefly, such a nonlinear process is based on the introduction of an intense external electric field, which is analogous to the laser intensity that induces nonlinear changes in both the electronic and atomic polarization of the materials composing the system under study. Thereby, the refractive index gains a quadratic dependence on the electric field strength, a.k.a., Kerr effect[76]. In this case, the nonlinear permittivity is given by the following relationship:

$$\varepsilon_r = \varepsilon_L + \chi^{(3)} |E_{loc}|^2, \qquad (2)$$

where $\varepsilon_r$ is the linear permittivity of the dielectric material, $\chi^{(3)}$ is the third-order nonlinear susceptibility, and $E_{loc}$ is the strength of the localized electric field at the interface. The magnitude of the Kerr effect depends on various factors including the material and structural properties, such as electronic configuration, optical transparency, and symmetry[77]-[78]. The breaking of Lorentz reciprocity can be accomplished by combining large optical nonlinearity with geometrical asymmetry, hence, achieving nonreciprocity without the need of an external bias (for example introducing strain, applying drift currents, or electric and magnetic fields).

To the best of our knowledge, self-induced nonreciprocal light transmission based on nonlinear topological photonic structures has not yet been investigated.

To demonstrate the self-induced nonreciprocal light transmission through the presented topological interface design, linear polarized light propagates forward (from left to right / blue color arrow in **Figure 7**(a)) and backward (from right to left / green color arrow in **Figure 7**(a)). Defects are introduced at the interface to make the topological photonic structure asymmetric, as can be seen in **Figure 7**(a). More specifically, we introduce two types of defects: (i) air defects within the dielectric section with radius $R_{Air}$ and (ii) dielectric defects within the air section of the checkerboard photonic crystal with radius $R_{Diel}$ both placed at the interface. The induced edge mode frequency is different for forward and backward propagation as the input power is increased in the case of defect dimensions $R_{Air} = 45$ nm and $R_{Diel} = 5$ nm with result shown in **Figure 7**(b). The induced contrast in transmission resonance frequencies in the case of backward and forward light propagation when the power is increased proves that the presented nonlinear system is nonreciprocal with relevant results depicted by the red line in **Figure 7**(b). It is important to note that Kerr nonlinearity is a relatively weak nonlinear mechanism and, as a result, is always treated in the perturbative regime during our nonlinear modeling. Therefore, changes induced by the Kerr effect cannot exceed in amplitude the linear (in the absence of laser power) refractive index. For that reason, we set the maximum allowable laser power value to $1.32 \ GW/cm^2$ (refer to **Figure 7**(b)). Note that such laser intensity values (and much higher spanning to hundreds of GW/cm$^2$) can be achieved with pulsed femtosecond lasers, as presented in various experimental papers[79],[80]. Additionally, lossless dielectric material with a third-order nonlinear susceptibility ($\chi^{(3)}$) of $2.8 \times 10^{-18} \ m^2/V^2$[76] was used, similar to silicon operating at near-IR frequencies[76]. Interestingly, the nonlinear response time of the presented self-induced nonreciprocity is due to the Kerr nonlinear effect that is based on the third-order nonlinear susceptibility of the material. Hence, it should have instantaneous or extremely fast response time[76] only limited by the Ohmic losses of the system, as was also shown in the relevant but non-topological self-induced photonic system experiments[57].

To further enhance and tune the contrast amplitude of the realized nonreciprocal response, we vary the dimensions of the defect-like structures introduced at the interface. Therefore, we also demonstrate that defects offer an additional control over the spectral positioning of the induced edge state. The defects are cylindrical, as shown in **Figure 7**(a), with a maximum radius of 55 nm, gradually reducing to 0 nm (i.e., absence of defect). To extract the self-induced nonreciprocity, we employ a similar approach: light propagation through the interface along either forward or backward direction illumination followed by computing the transmission spectra. The difference between the induced edge mode frequency for backward and forward light propagation ($\Delta f = f_{Fwd} - f_{Back}$) provides a nonreciprocity contrast metric (see **Figure 7**(c)). Depending upon the radius of both dielectric and air defects, we observe a significant change in this nonreciprocity metric. To lay this out clearly, we perform a series of nonlinear simulations, where we fix the value of the input intensity to 1.09 $GW/cm^2$ and vary the radius of both defects from 0 to 55 nm. The resulting nonreciprocity contrast metric data is presented as a color density plot in **Figure 7**(c). Interestingly, not all arbitrarily chosen defect sizes can introduce nonreciprocity. We found the following relationship between the air defect radius ($R_{Air}$) and dielectric defect radius ($R_{Diel}$): $R_{Air} = -0.0141 R_{Diel}^2 + 1.3998 R_{Diel} + 1.1751$, which leads to a reciprocal arch-kind gap curve for the specific input intensity (1.09 GW/cm$^2$). Note that this behavior will have another shape for different input intensity values. Moreover, we observe that the nonreciprocity intensifies when the difference between the radius of the air defect compared to the dielectric defect is more pronounced, meaning larger geometrical asymmetry. Interestingly, the strongest nonreciprocity with negative amplitude is obtained when the air defect radius is $R_{Air} = 55$ nm and the dielectric defect is absent ($R_{Diel} = $ 0). The strongest positive nonreciprocity is obtained at the inverse scenario when the dielectric defect radius is $R_{Diel} = 55$ nm and the air defect is absent ($R_{Air} = 0$). These results can be explained as follows: the dielectric material addition to the system results to a spectral red shift of the edge mode resonance, while dielectric material removal (i.e., the introduction of air defect) leads to a spectral blue shift, as also depicted in **Figure 6**(a). In order to demonstrate the induced different edge state frequencies obtained under forward and backward illumination, the electric field distributions are presented for these frequencies. An induced edge state is obtained only for forward illumination and not backward illumination or vice versa, as shown in **Figure 7**(d) and **Figure 7**(e). In these plots, we clearly derive that the confinement of light at the interface occurs only at the edge state frequency that corresponds to the appropriate illumination direction.

**Topologically Protected Nonreciprocal Response**

It is worth noting that asymmetric photonic crystal systems can achieve self-induced nonreciprocal response[38],[39],[52] but without the topological protection presented in our work. To prove that nonreciprocity and topological protection coexist in the current designs, we explore a step-shaped topological photonic interface with two sharp corners (similar to **Figure 4**(c)), however, when defects are introduced along the interface to realize geometric asymmetry leading to self-induced nonreciprocity. This type of configuration will not become nonreciprocal if a usual asymmetric photonic crystal waveguide is used[38],[39],[52] because topological protection does not occur in such waveguides, and the edge state ceases to exist mainly due to leakage and scattering from the abrupt corners.

Nevertheless, in the current topological system, we achieve self-induced nonreciprocity even in this sharp corner design, as demonstrated in **Figure 8**. More specifically, the induced edge states under forward and backward illumination in the case of checkerboard structures ($a = 137$ nm) with defects ($R_{Diel} = 5\ nm$ and $R_{Air} = 45\ nm$) throughout the interface are shown in **Figure 8**(a) and **Figure 8**(b). Moreover, the spectral evolution of the transmission coefficient for the two illumination cases is plotted in **Figure 8**(c) where it is clearly proven that we realize self-induced nonreciprocal response despite the abrupt corners due to the topological protection of the induced edge mode. Note that high input intensity radiation with equal value is used for both illumination cases to trigger the nonlinear Kerr effect leading to the presented self-induced nonreciprocity.

It is worth mentioning that the magnitude of the nonreciprocal response (e.g., frequency shift difference and transmission contrast) relies on the applied incident power and the size of perturbation defects placed at the topological interface, as was demonstrated before in **Figure 7**(b) and **Figure 7**(c). However, we clearly demonstrate in **Figure 8**, that the edge mode is topologically protected even in the case of self-induced nonreciprocity. Thus, the existence and continuity of the edge state itself is always topologically protected despite the frequency shift or transmission contrast due to nonreciprocity. Note that we always confirm the topological protection of the induced nonreciprocal edge states, even in the case of nonlinear response, by performing the Zak phase analysis, as depicted in **Figure 3**. Hence, it can be concluded that the edge mode is topologically protected even in the nonreciprocal case.

**Realistic Design Operating in Visible Frequencies**

The effect of loss on the topological photonic resonance mode is investigated in detail in Supplementary Material Section 1. Based on our simulations, the edge state is found to be sensitive to loss in addition to the value of the used material refractive index. More specifically, we need to use a dielectric material with a high refractive index that enhances the contrast between air and dielectric material sections in our proposed structures leading to stronger local optical density. In Supplementary Material Sections 2-3, the performance of two different realistic dielectric materials: Si and titanium dioxide (TiO$_2$) are investigated, where we prove that the presented nonreciprocal system can work for such realistic materials at near-IR and ultraviolet (UV) frequencies.

Here, we focus our attention to the technologically important visible frequency range and demonstrate nonreciprocal topological performance by using photonic crystals made of TiO$_2$ where the material losses are taken into full account. This material (TiO$_2$) has low losses in the visible range combined with a third-order nonlinear susceptibility ($\chi^{(3)}$) of $2.1 \times 10^{-20} \frac{m^2}{V^2}$ [76]. In the current study, photonic crystals have dimensions of $24 \times 1$ (12 unit cells of PhC-1 and 12 unit cells of PhC-2) unit cells with a lattice parameter of 153 nm, where the air hole and TiO$_2$ pillar defects at the interface are introduced with approximately radii of 50 nm and 6 nm, respectively. **Figure 9(a)** clearly shows the presence of an edge state (red colored line) within the bulk energy band diagram of the realized TiO$_2$ based topological photonic crystal interface (structure design is depicted in **Figure 1**(e)). It is also computed that the topological edge state frequency dependence to the lattice constant $a$ is given by $f_{edge} = \beta/a$, where the constant is $\beta = 1.343 \times 10^5 \, THz.nm$.

In the case of lattice constant equal to 153 nm, we successfully engineer a narrow transmission leakage mode in the visible range ($\approx$ 557 THz) for forward and backward illumination, as depicted in **Figure 9**(b) and (c), where the effect of power intensity on the nonreciprocal behavior of the induced topological edge state is also demonstrated. We observe that TiO$_2$ based photonic crystals can further boost the nonreciprocity frequency contrast to ~ 0.08 THz at laser intensity of 1.32 $GW/cm^2$ (see **Figure 9**(d)).

One notable limitation of Kerr nonlinearity-based nonreciprocal devices is dynamic reciprocity[81]. This limitation emerges when small optical signals or noise interact with large incident signals traveling in the opposite direction. To demonstrate the dynamic reciprocity effect, our nonlinear checkerboard photonic design with circular shape defects (TiO$_2$ and air) located at the interface is simultaneously excited with input waves from both sides (forward and backward). A schematic illustrating this configuration is shown in the inset of **Figure 10**. The maximum transmission contrast was achieved at 557.3 THz (see **Figure 9**(c)) when $I_{inF} = 1\ GW/cm^2$, where $I_{inF}$ is the incoming forward intensity that has a fixed high value. Next, we sweep the backward input intensity, $I_{inB}$, from low values of $6.54 \times 10^{-4}\ MW/cm^2$ to $6.21\ MW/cm^2$. The output power was probed from both sides by using the integral of $\int \int_C \vec{S} \cdot \vec{n}$ where $\vec{S}$ is the pointing vector crossing the upper boundary curve C and $\vec{n}$ is the boundary norm vector [82]. Since our simulations are performed in the frequency domain, the measured output power at each port represents the sum of both the excited and transmitted power to that probe. To obtain the actual output power values ($P_{outF}$ and $P_{outB}$), we extract the input power values ($P_{inF}$ and $P_{inB}$) from the total power measured at each port ($P_{inF} + P_{outB}$ and $P_{inB} + P_{outF}$), respectively. Output power ratios at each side are calculated by dividing the output power values by the sum of input power values, and these ratios are demonstrated in **Figure 10** as a function of the excitation from backward with varying intensity of $I_{inB}$. The output power ratios at the forward and backward ports gradually converge as $I_{inB}$ values increase. Interestingly, approximately at $I_{inB} = 0.6\ MW/cm^2$, the normalized output power ratios become nearly identical, indicating the onset of dynamic reciprocity limitation. Counterintuitively, as shown in **Figure 10**, nonreciprocal transmission can still be maintained provided that the input intensity from the opposite (backward) direction remains below this threshold value. Importantly, despite this limitation, the proposed passive photonic crystal design shows nonreciprocity over a relatively broad range of input intensities. This characteristic indeed distinguishes our design from other relevant works that rely on external biasing (e.g., magnetic field [45]-[48], [83]) to tune the nonreciprocal response.

## Comparison with Previous Works

Another important parameter in the development of the presented nonreciprocal PTIs is the transmission contrast at the minimum insertion loss as it defines the degree of unidirectional behavior while minimizing input signal attenuation, ensuring both nonreciprocal efficiency and performance in integrated photonic circuits[55],[84]-[85]. Reporting the transmission contrast at the point of optimal (minimum) insertion loss offers a consistent and informative benchmark for clear assessment of the performance of various nonreciprocal photonic device designs.

This section is devoted to reporting the transmission contrast performance of our proposed PTI design based on two different materials systems: (i) $TiO_2$-air and (ii) Si-air and comparing their performance with previously published relevant designs. We observe that the material type, operation spectra, and defect incorporation are among the most important factors that affect the nonreciprocal transmission contrast. In the previously presented designs, the $TiO_2$-based photonic structure in **Figure 9** has a maximum transmission of 45% in the visible spectrum range, and the Si-based photonic structure in Figure S2 has a maximum transmission of 24% in the NIR spectral region. Here we optimize these two designs with the goal of minimizing loss from material extinction (**Figure S1**). This happens by aligning their operational spectrum with regions of near-zero material extinction coefficient by tuning the lattice parameter of the photonic crystal. Reducing the size of air defects placed at the interface also enhances transmissivity but decreases nonreciprocity. Therefore, precise engineering and fine-tuning of these structural parameters are crucial for achieving optimized high transmission contrast while preserving strong nonreciprocity.

Note that the optimized reciprocal PhC structure design without defects achieves a maximum transmission of 100% (see **Figures 2(c) and (d)**). To convert this reciprocal system into a nonreciprocal photonic crystal structure, defects at the interface should be introduced, as explained in previous sections. Large air and dielectric defect sizes in a photonic crystal system lead to higher nonreciprocity and lower maximum transmission values. To achieve higher nonreciprocal transmission contrast, defect sizes should be optimized. Reducing the air defect size in the PTI of **Figure 9** (TiO$_2$-air system) by approximately 50% increases its maximum transmission to 96%, as shown in **Figure 11(a)** and maintains the nonreciprocity frequency bandwidth to around 12.5 GHz. Likewise, a 60% reduction in the air defect size (of the same PTI Si-air system shown in **Figure S2**) results in a maximum transmission of 97% (see **Figure 11(b)**). These responses are achieved in both cases for a fixed maximum power of 1.32 GW/cm$^2$ which is much lower than the laser damage threshold of the presented systems. More specifically, the laser damage threshold value in any nonlinear optical system will exhibit variations depending on several factors including laser wavelength and pulse duration. For the currently proposed photonic crystal system, the used maximum intensity value is below the laser damage threshold value for both Si (0.18–0.27 J/cm$^2$ [86]-[90]) and TiO$_2$ (4.69-9.0 J/cm$^2$ [91]-[93]).

As summarized in Table 1, our design demonstrates significantly improved performance compared to previously reported experimental Kerr-type nonreciprocal systems and comparable or improved metrics to the theoretical estimations of other relevant structures. However, more importantly, our all-dielectric planar PTI system achieves high figures of merit without relying on active gain materials to boost transmission[55], [57], [94]-[95], which are known to introduce instability limitations and other practical challenges. Therefore, our passive (gain-free) architecture presents a robust and practical alternative for realizing high-performance topological protected nonreciprocal photonic systems.

## Conclusions

To conclude, we demonstrate a narrowband topological leaky edge mode induced at the interface between two photonic crystal designs based on all-dielectric complementary checkerboard structures. Despite identical energy dispersion diagrams on both sides of the interface, differences in Zak phases are calculated, proving the generation of a topological edge state. The interaction of circularly polarized light with the designed interface exhibits unidirectional propagation of light which is indicative of a topological edge mode possessing spin locking. Moreover, we present enhanced and tunable self-induced nonreciprocity performance by incorporating the nonlinear Kerr effect in conjunction to geometrical asymmetry when defects are introduced along the topological interface. Therefore, our findings provide valuable insights into the emerging field of nonlinear topological photonics that can open new avenues for the potential development of next-generation photonic devices and quantum information systems[60].

## Acknowledgements

This work was partially supported by the National Science Foundation (NSF) under award numbers ECCS 2329940, DMR 1808715, CMMI 2211858, and OIA-2044049 Emergent Quantum Materials and Technologies (EQUATE), Air Force Office of Scientific Research under award numbers FA9550-18-1-0360, FA9550-19-S-0003, FA9550-21-1-0259, and FA9550-23-1-0574 DEF, Swedish Knut and Alice Wallenbergs Foundation supporting grant titled 'Wide-bandgap semi-conductors for next generation quantum components', and American Chemical Society/Petrol Research Fund. C. A. and E. S. acknowledge partial support by NSF 2224456. M. S. acknowledges the University of Nebraska Foundation and the J. A. Woollam Foundation for financial support.

## Conflict of interest:
The authors declare that they have no competing interests.

## Data and materials availability:
All data needed to evaluate the conclusions in the paper are present in the paper and/or the Supplementary Materials.

# Figures

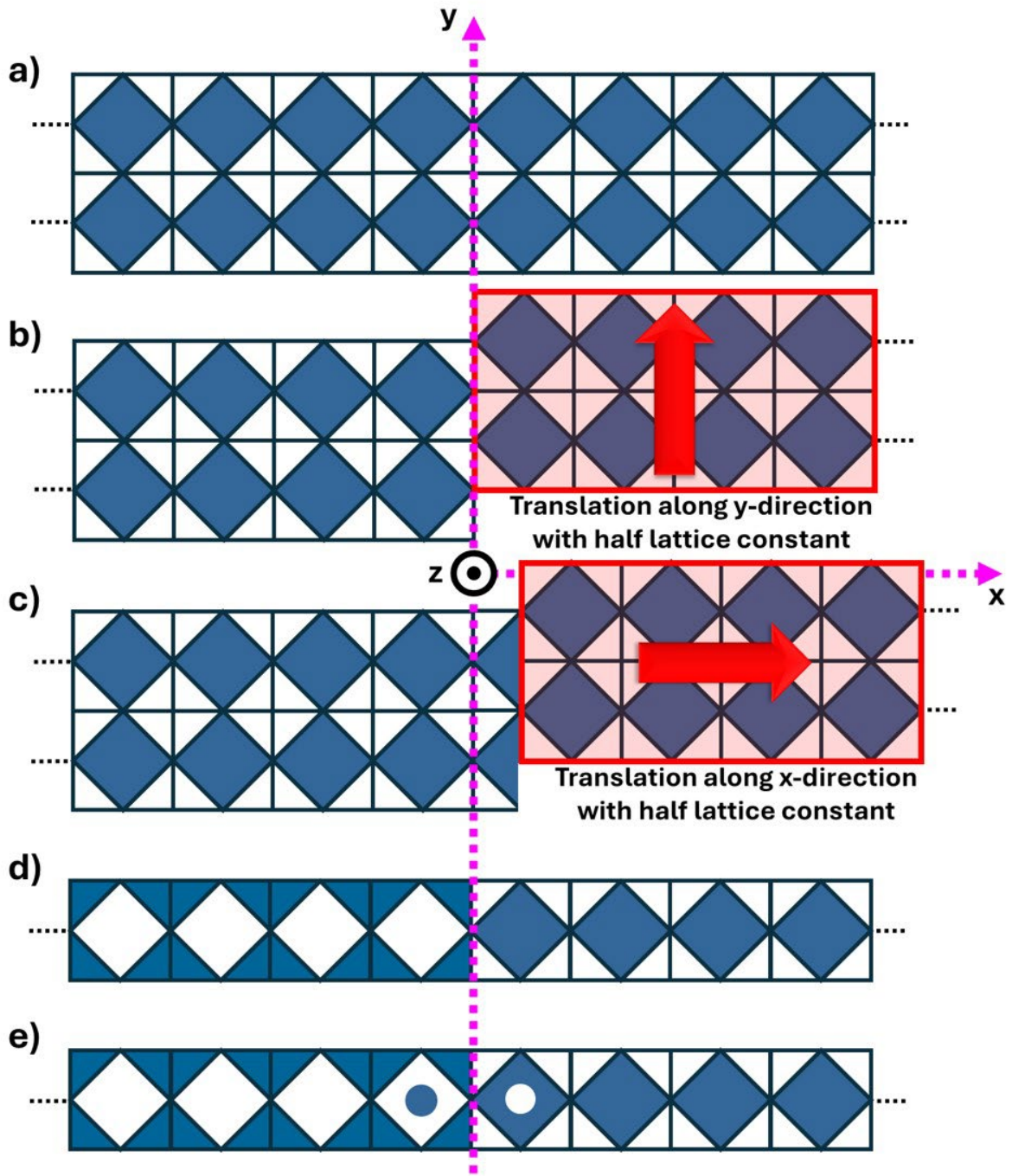

**Figure 1** (a) Proposed 2D photonic crystal design made of an array of all-dielectric checkerboard unit cell structures. (b)-(c) By shifting half of the checkerboard photonic crystal towards both (b) y- and (c) x-directions by half unit cell lattice constant $a/2$, we create a topological edge state along the realized interface. (d) Schematic representation of the semi-infinite photonic crystal structures when there is a topological interface. (e) Introduction of dielectric pillar and air hole defects at the interface to realize asymmetry in the geometry leading to self-induced nonreciprocity due to the inherent nonlinear response.

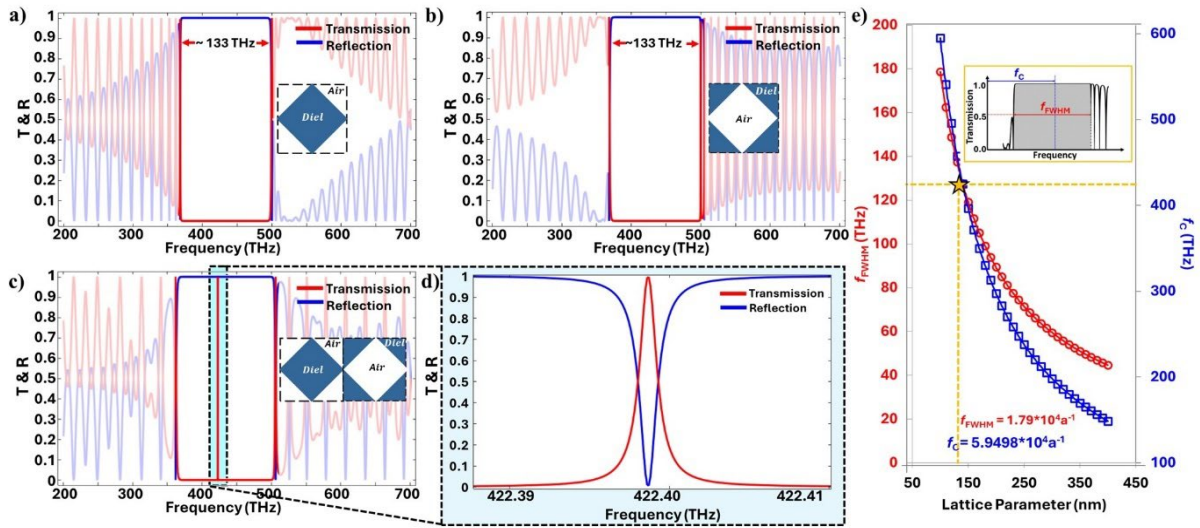

**Figure 2** (a)-(b) Computed transmission (red solid line) and reflection (blue solid line) spectra of (a) PhC-1 and (b) PhC-2. (c) Computed transmission (red solid line) and reflection (blue solid line) spectra when an interface is formed between PhC-1 and PhC-2. (d) Zoom in photonic band gap region shown in Fig. 2(c) to highlight the emergence of a narrow leaky resonant mode due to the topological interface formed between PhC-1 and PhC-2. (e) The evolution of both full-width half-maxima ($f_{FWHM}$) and central ($f_c$) frequencies of photonic band gap as a function of the lattice parameter $a$. The inset is the corresponding spectral evolution of the transmission coefficient for a lattice constant of 140 nm (star point in the figure) and the gray hatched area is the transmission spectra depicting the induced photonic band gap.

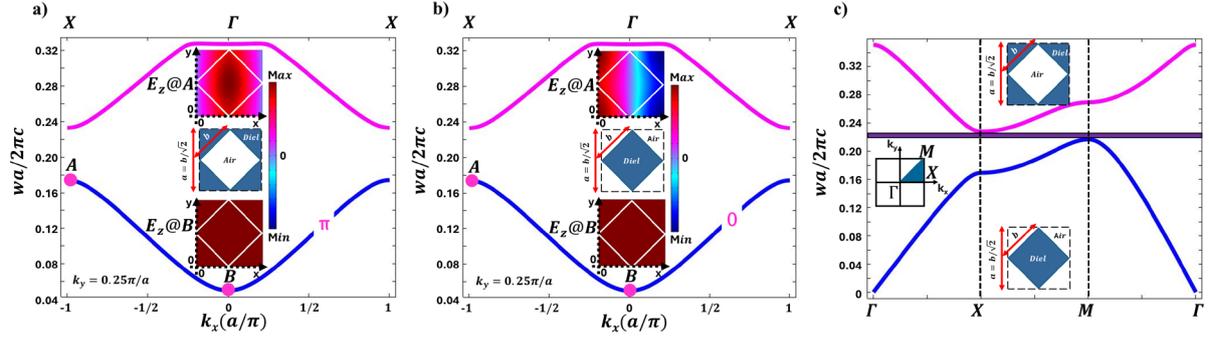

**Figure 3** (a) Unit cell dispersion diagram for PhC-1 which has air inside and dielectric material around the checkerboard design. The inset plots demonstrate the $E_z$ at two high symmetry points depicted as A and B. When $|E_A| = 0$ and $|E_B| \neq 0$ at x=0, the Zak phase is $\pi$. (b) Unit cell dispersion diagram for PhC-2 which has dielectric material inside and air around the checkerboard design. The inset plots demonstrate the $E_z$ at two high symmetry points depicted as A and B. When $|E_A| \neq 0$ and $|E_B| \neq 0$ at x=0, the Zak phase is 0. (c) Complete dispersion band diagram for the PhC-1 and PhC-2 unit cells. The left inset shows the first Brillouin zone. The formation of a band gap ranging from $0.216 c/a$ to $0.227 c/a$ is demonstrated by the purple region. This band gap is shared by PhC-1 and PhC-2 due to the self-complementary of the checkerboard PhC design.

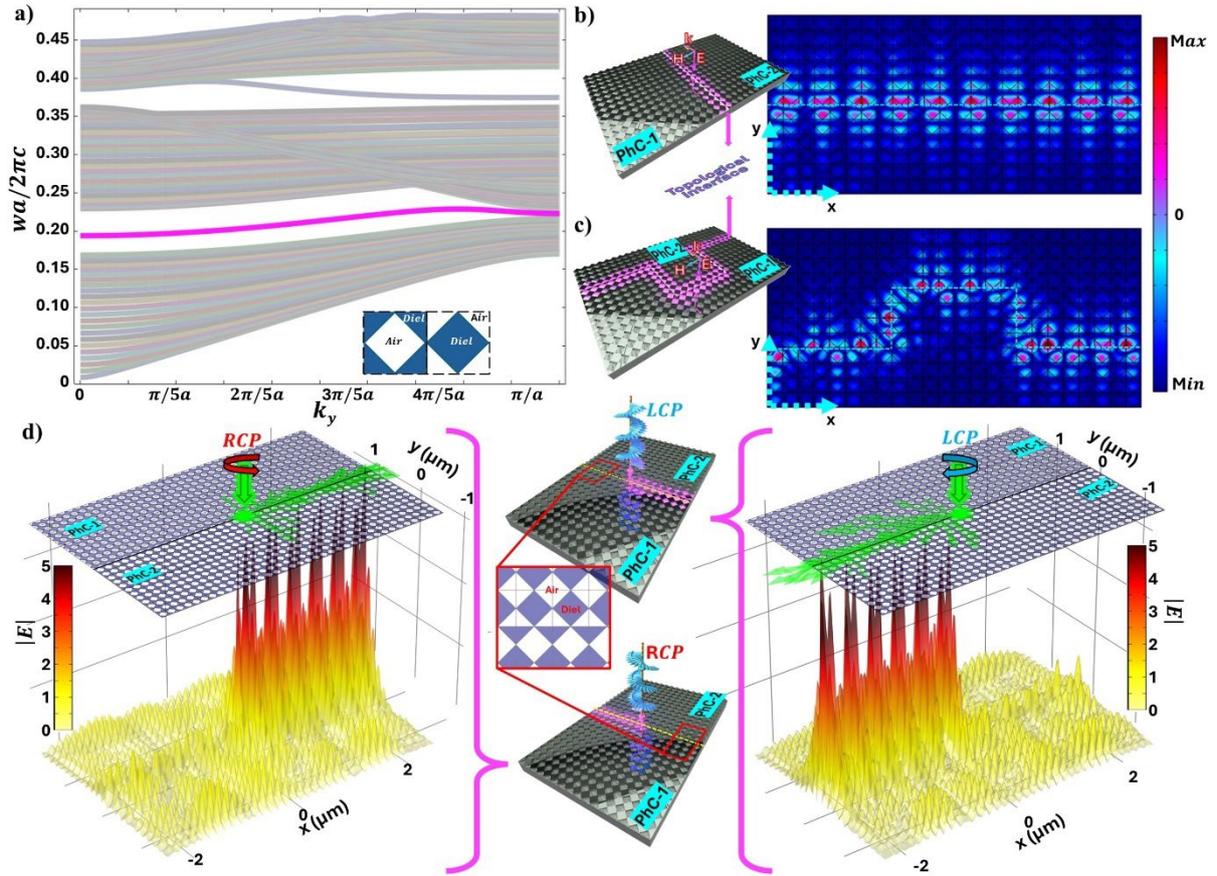

**Figure 4** (a) Band diagram of the photonic boundary formed between PhC-1 and PhC-2. The magenta line shows the edge state dispersion. (b)-(c) The coupling of incident electromagnetic waves at two different interface scenarios: (b) straight line and (c) step-shaped line with two sharp corners. (d) Directional chiral propagation of circularly polarized light at the interface, a.k.a., spin locking. While RCP light propagates to the +x direction, the LCP light propagates to the opposite direction.

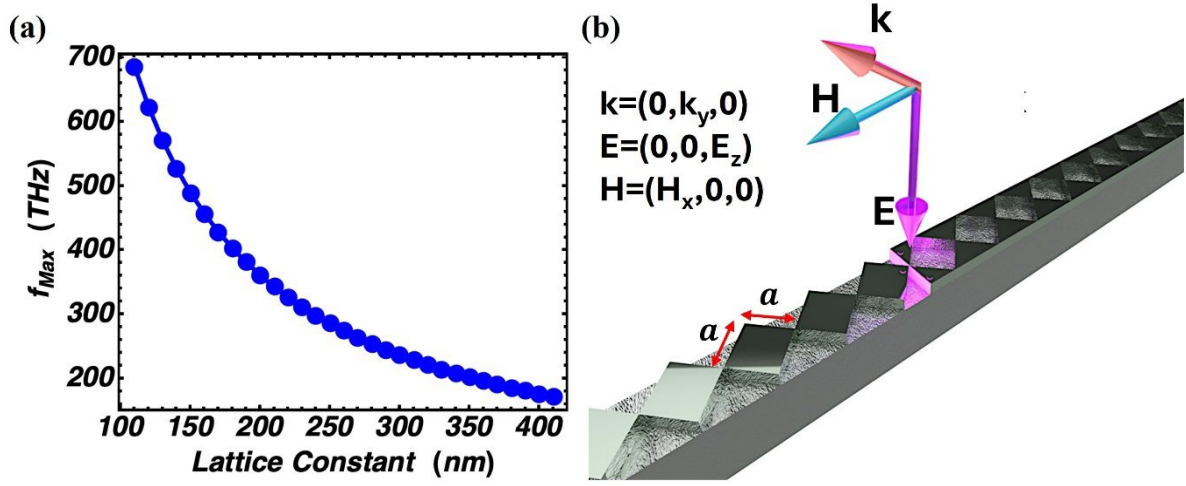

**Figure 5** (a) Frequency shift in the induced edge mode transmission peak as the lattice constant $a$ varies. (b) Schematic representation of the single row topological photonic structure which depicts the directions of electric (E) and magnetic (H) fields, in addition to the propagation vector (k).

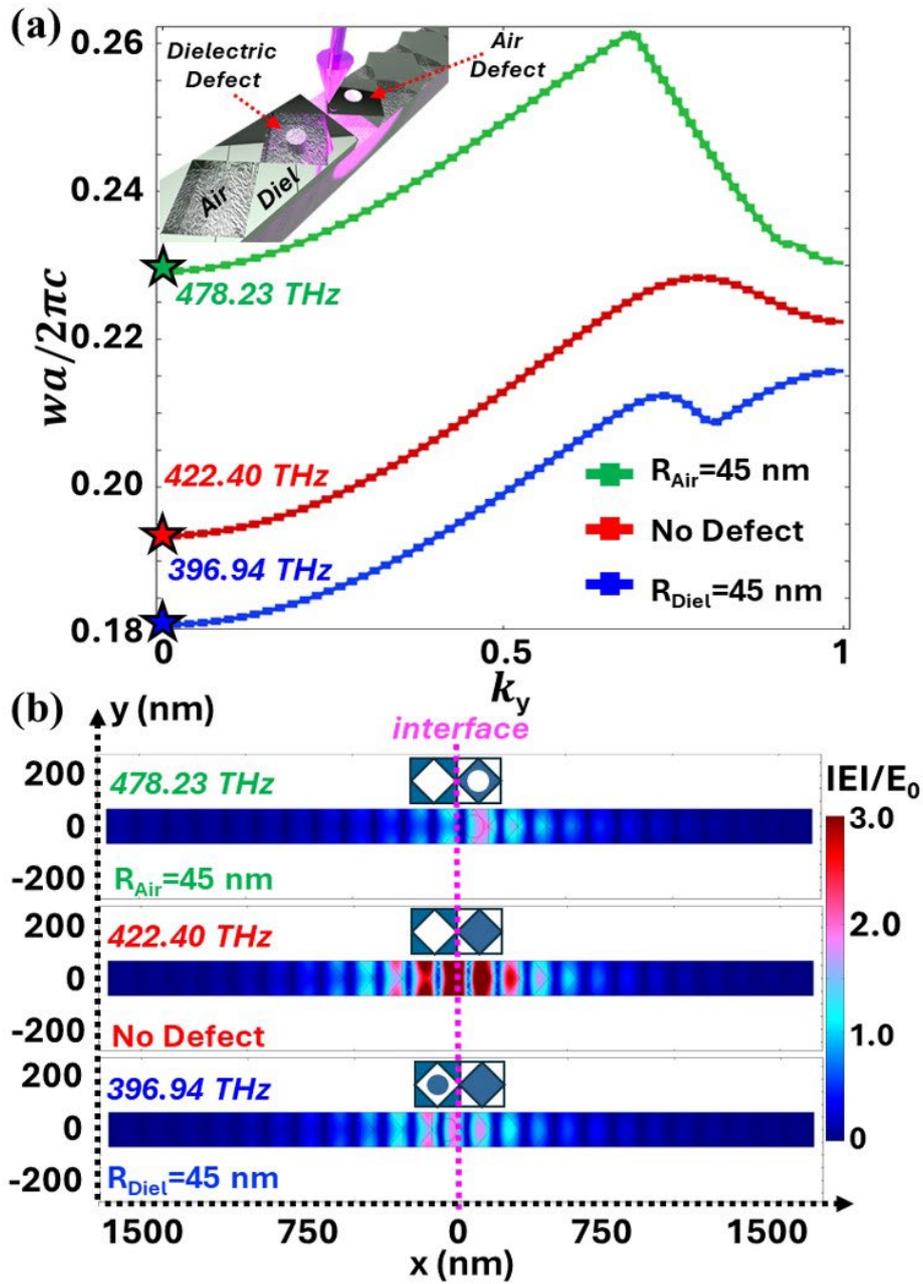

**Figure 6** (a) Eigenfrequency simulation results demonstrating the evolution of the topological edge mode both in the presence of air hole (green square symbols) or dielectric pillar (blue square symbols) defects, and in the absence of defects (red square symbols). (b) The corresponding normalized electric field distributions at k = 0 for air hole defects (top), no defects (middle), and dielectric pillar defects (bottom).

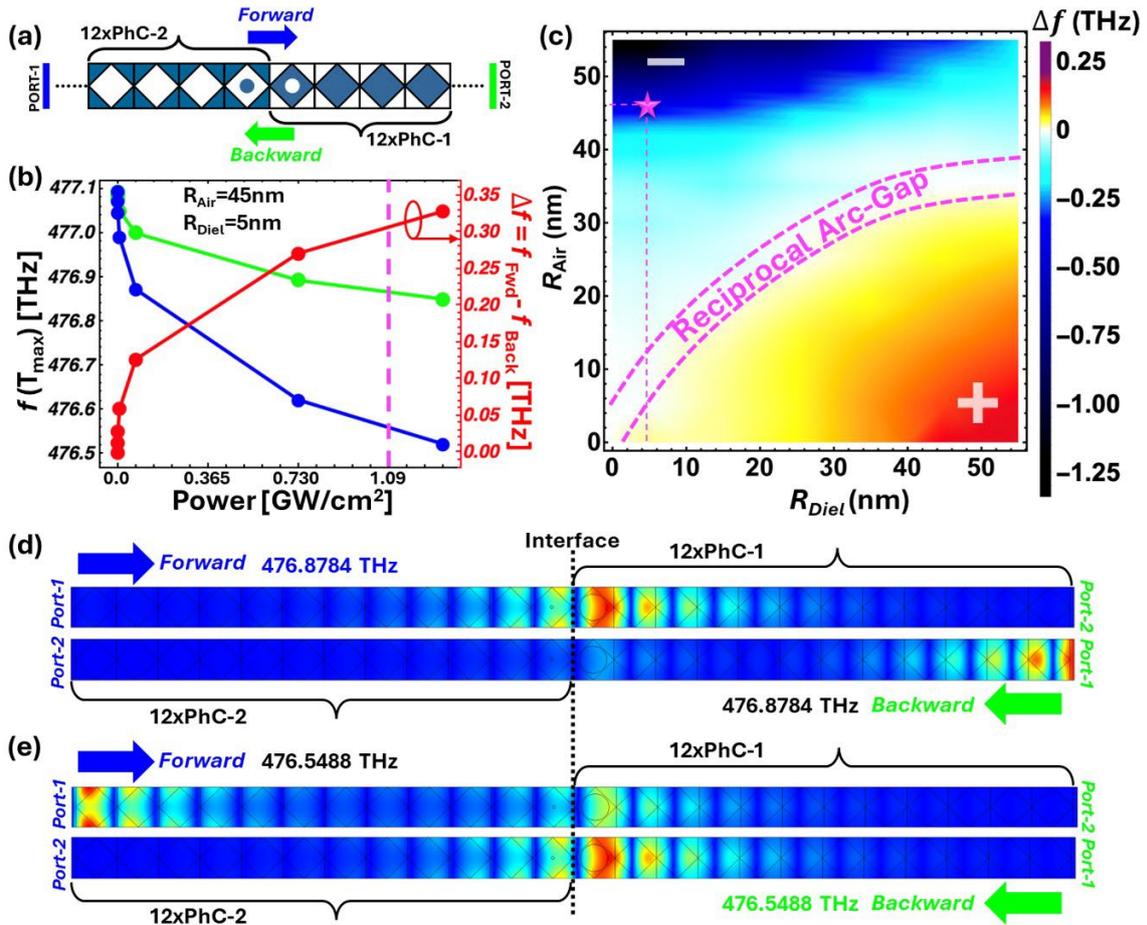

**Figure 7** (a) Schematic diagram of the asymmetric topological photonic crystal design that is used to achieve self-induced nonreciprocity based on the Kerr nonlinear effect. (b) Power-dependent evolution of topological edge mode transmission extrema frequency values under forward (blue) and backward (green) light illumination. The nonreciprocity contrast is quantified as the difference between these frequencies: $\Delta f = f_{Fwd} - f_{Back}$ (red). (c) Bisignate nonreciprocity contrast as a function of defect size. (d)-(e) Nonreciprocal edge modes induced in the case of forward ($f_{forward}$=476.8784 THz, (d)) and backward ($f_{backward}$=476.5488 THz, (e)) illumination.

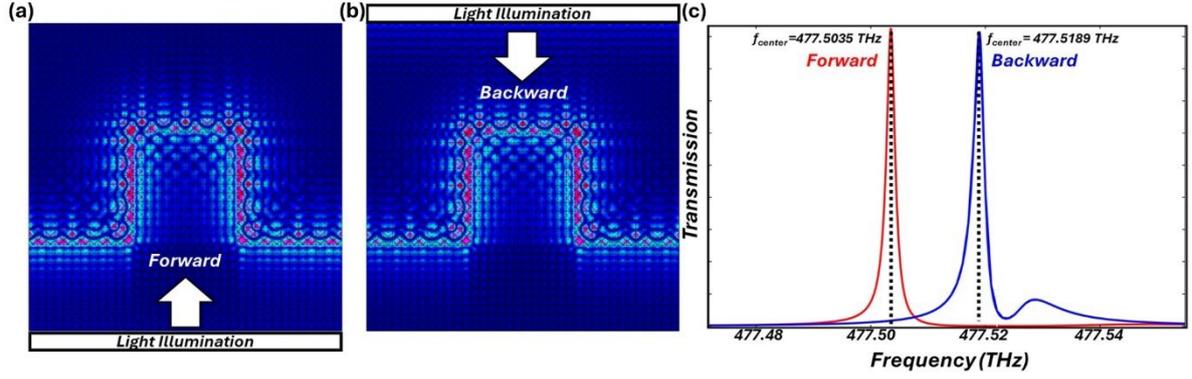

**Figure 8** (a) Forward and (b) backward illumination of a step-shaped topological photonic interface with two sharp corners where defects are introduced along the interface to realize geometric asymmetry leading to self-induced nonreciprocity. (c) Different forward and backward transmission spectra due to Kerr nonlinear effect leading to self-induced nonreciprocal response with topological protection.

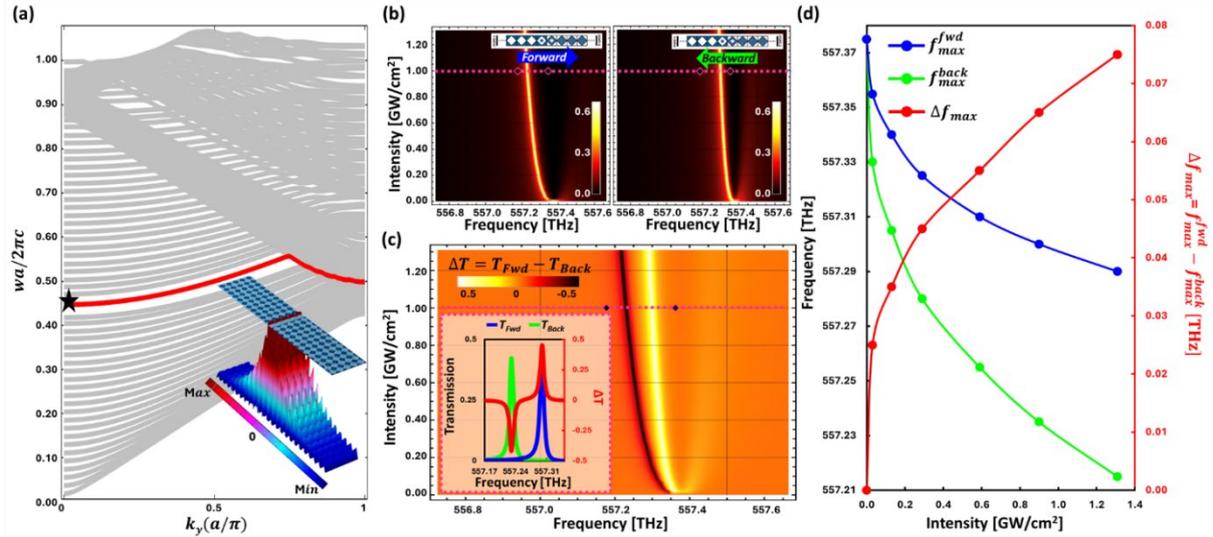

**Figure 9** (a) Dispersion diagram of photonic topological structure made of $TiO_2$ and air defects (air hole and Si pillar with radii of 50 nm and 6 nm, respectively). (b) The edge mode peak experiences a red shift by changing intensity values in the case of forward (left, $T_{Fwd}$) and backward (right, $T_{Back}$) illumination. (c) Nonreciprocal transmission contrast ($\Delta T = T_{Fwd} - T_{Back}$) is calculated and demonstrated as a function of intensity and frequency revealing the self-induced nonreciprocity in the photonic topological system. The inset figure shows the spectra of the edge mode transmission peak for forward (blue solid line) and backward (green solid line) illumination and their difference (red solid line) when the laser intensity is 1 GW/cm². (d) The trace of maximum resonant frequency difference derived by both color density plots shown in (b) for forward (blue color) and backward (green color) illumination cases. The trace of maxima nonreciprocal transmission contrast is also depicted in the same plot (red color).

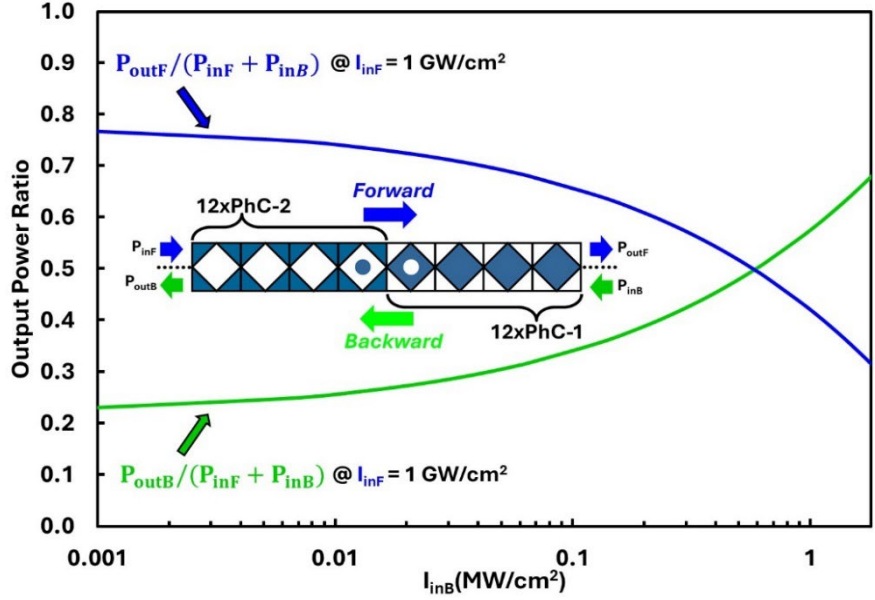

**Figure 10** Output power ratios at each side as a function of the backward excitation with varying intensity of $I_{inB}$ and fixed forward excitation with intensity $I_{inF} = 1\text{GW/cm}^2$. The inset schematic representation shows the photonic crystal design with defects when it is simultaneously excited by two input signals from opposite directions.

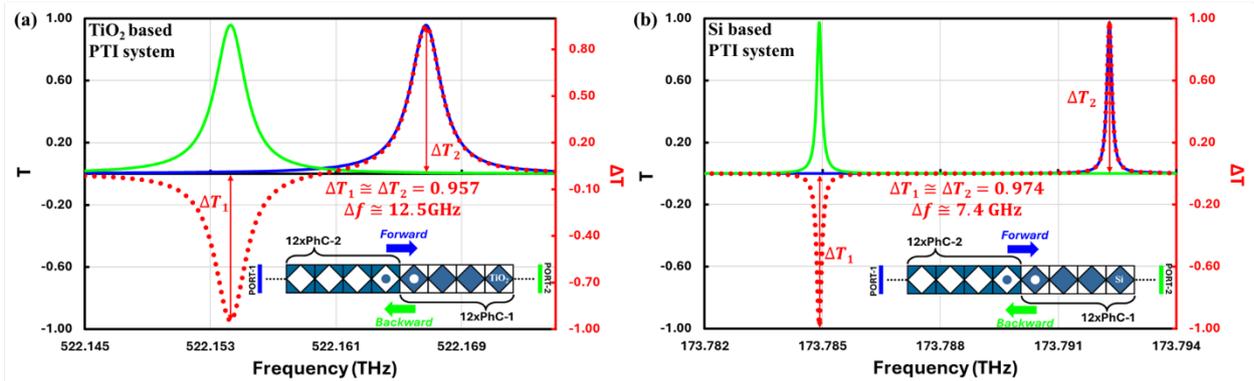

**Figure 11** (a) Backward (green) and forward (blue) transmission peaks (topological edge mode) of the $TiO_2$-air based PTI system for a fixed maximum power of $1.32\text{GW/cm}^2$. The red dashed line shows the transmission difference between forward and backward transmission peaks. (b) Backward (green) and forward (blue) transmission peaks (topological edge mode) of the Si-air based PTI system for a fixed maximum power of $1.32\text{GW/cm}^2$. The red dashed line shows the transmission difference between forward and backward transmission peaks.

# Tables

Table 1: Comparison between different works that show nonlinearity-induced nonreciprocity

| Reference No. | Bandwidth /center frequency | Kerr nonlinear coefficient [$\chi(3)(m^2/V^2)$] | Thickness/wavelength | Signal Intensity | Transmission contrast at minimum Insertion Loss |
|---|---|---|---|---|---|
| [53] | | $2.8 \times 10^{-18}$ | (2.7-6.15) [μm] / 1.53 [μm] | (1.5-2) MW/cm$^2$ | -25.4/-35.7/-15.2 dB at -0.46/-0.41/-0.044 dB over NRIR of (2.79/1.5/1.52) dB(*) |
| [54] | | $2.79 \times 10^{-18}$ | 0.1 [μm] / 1.5 [μm] | 5 kW/cm$^2$ | -17 dB at -1.2 dB over 4.77 dB(*) |
| [55] | | | 540 [nm]/1.47 μm | 22.3 kW/cm$^2$ | ~ -7.2 dB at ~ -2.7 dB over NRIR of ~1.95 dB(*) |
| [56] | | $2.8 \times 10^{-18}$ | 100 [nm]/ 1.54 [μm] | 3 kW/cm$^2$ | The non-reciprocal ratio is 10.7 dB at insertion loss of 2.3 dB over NRIR of 2.6 dB(*) |
| [57] | | | 0.22 [μm]/ 1.55 [μm] | 4.55–9.05 dBm | The maximum transmission contrast is 20.3 at insertion loss of 1.1 dB over NRIR of 6.3 dB |
| [58] | 2.1 THz/349.4 THz | $6 \times 10^{-20}$ | 72 [nm] / 850 [nm] | 8 MW/cm$^2$ | -26.98 dB at -0.017 dB over NRIR of 3.1 dB |
| [96] | 0.6 THz/192 THz | | (1.33-5.334) [μm] / 1.56 [μm] | (16.8 ± 0.001) GW/cm$^2$ | -56 dB at -0.04 dB -65 dB at -0.2 dB |
| **This work (TiO$_2$)** | | $2.1 \times 10^{-20}$ | 153 [nm]/ 574 [nm] | 1.32 GW/cm$^2$ | -21.9 dB at -0.22 dB(*) |
| **This work (Si)** | | $2.8 \times 10^{-18}$ | 330 [nm]/ 1.7 [um] | 1.32 GW/cm$^2$ | -40.2 dB at -0.11 dB(*) |

(*) Transmission contrast at minimum Insertion Loss values is obtained for Ref. [53] from Figs. 3, 4, and 5, for Ref. [54] from Fig. 3(c), for Ref. [55] from Fig. 3(b), for Ref. [56] from Fig. 3(d), and for this work (TiO$_2$ and Si) from Fig. 11.

# Supplementary Material

# Self-induced nonreciprocity from asymmetric photonic topological insulators

Sema Guvenc Kilic[1,#], Ufuk Kilic[1], Mathias Schubert[1,2], Eva Schubert[1] and Christos Argyropoulos[3,*]

[1]Department of Electrical and Computer Engineering, University of Nebraska-Lincoln, Lincoln, NE 68588, USA

[2] Solid State Physics and NanoLund, Lund University, P.O. Box 118, 22100, Lund, Sweden

[3] Department of Electrical Engineering, The Pennsylvania State University, University Park, PA 16803, USA

[#]sguvenckilic2@huskers.unl.edu and *cfa5361@psu.edu

In the Supplemental Material section three key systematic studies are performed and discussed as listed below:

- The effect of loss on the resonance mode within the topological photonic gap.

- Near-IR nonreciprocal behavior of topological edge state from Si photonic crystal design.

- The nonreciprocal behavior of topological edge state from $TiO_2$ photonic crystal design.

The nonreciprocal behavior of the topological edge state relies on both the structural parameters and the material's linear and nonlinear optical properties. In this part, we investigate the incorporation of material loss which is expected to affect the narrowband resonance transmission spectra shown before in Figure 2(d). It will broaden the full-width half maxima (FWHM) of the resonance and, therefore, decrease its quality factor, as depicted in Figure S1(a). The inclusion of material loss also leads to a spectral shift in the resonance mode (see Figure S1(b)).

Since the loss of Si material takes nonzero values above $f \approx 195\ THz$, instead of having the topological edge state active in the visible range, we can achieve a strong resonance mode in the near-IR part of the spectrum. Thereby, the lattice constant, air hole, and Si pillar defects are taken as 330 nm, 108 nm, and 12 nm, respectively. Both air hole and Si pillar defects are employed because their presence enhances the nonreciprocal performance of the topological edge state (see Figure 7). The nonreciprocal transmission performance as a function of incident power intensity of the resulting Si-based topological insulator design is summarized in Supplementary Material Figure S2. Regarding the nonreciprocal performance of topological edge state from Si-based photonic crystal design, we observe that the onset of absorption limits the use of Si in topological edge state creation to the near-IR spectral range or larger wavelengths. In this spectral range (i.e., near-IR), we can achieve strong nonreciprocity that is measured to be 0.134 THz for the laser intensity of 1.32 $GW/cm^2$ (see Supplementary Material Figure S2).

1. **Effect of loss on the resonance mode within the topological photonic gap**

In the first part of our paper, we show the response of our photonic crystal designs from an artificial material with a refractive index with value ~3.4 (i.e., permittivity equal to 11.56) and zero extinction coefficient, mimicking silicon at near-IR wavelengths. Here, we show how loss affects the induced topologically protected edge states by employing the dielectric function of Si material with and without losses. We utilize the spectroscopic ellipsometry-based measured complex dielectric function of crystalline state Si reported by D. Franta et. al. [1]. We observe a gradual increase in the extinction coefficient beginning around f = 250 THz (λ = 1200 nm). Moreover, due to the significant loss factor present in the visible and UV sections of Si dielectric function, the total absorbance of the overall periodic photonic crystal platform results in the disappearance of both field localization effect at the topological interface and the transmissive narrowband leakage mode in the photonic gap. This effect can be clearly seen in Figure S1(a) which shows the transmission intensity spectra as a function of lattice constant. We observe that the lattice constant (*a*: varied from 400 nm to 500 nm) and the corresponding central frequency (*f*) of the resonance mode are inversely proportional to each other. The increase in the lattice constant leads to a shift to the central resonant frequency into the near-IR part of the spectrum (*f*: varied from 310 THz to 249 THz) where silicon losses are almost zero. Thereby, we observe that the transmission intensity reaches its maxima value (one) when the lattice constant is 500 nm and the extinction coefficient is approximately ~ $10^{-7}$ (Figure S1(a)). Figure S1(b) shows both traces of transmission resonance maxima and their corresponding FWHM as a function of the lattice parameter. Figure S1(b) also demonstrates the corresponding spectral location of the transmission resonance maxima.

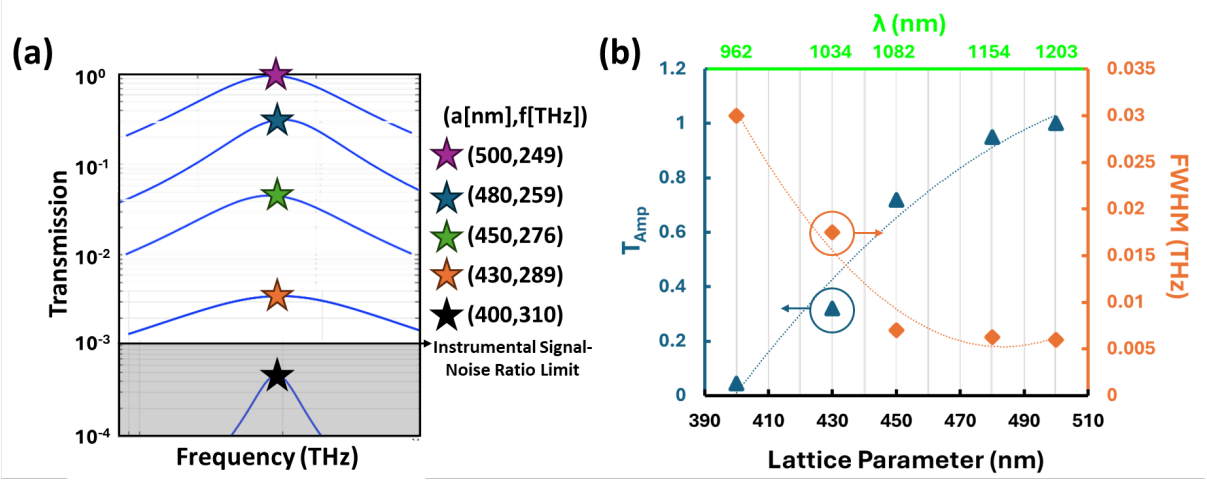

**Figure S1** (a) Spectral evolution of transmission coefficient in logarithmic scale for different lattice constants ranging from 400 nm to 500 nm when the realistic dielectric function of Si material is included. It is demonstrated that the resonance mode transmission diminishes as the frequency is increased. Moreover, the edge state central frequency shifts into a higher value when the lattice constant value is reduced. (b) The traces of FWHM and the central frequency as a function of the lattice parameter obtained from the transmission spectrum.

## 2. Near-IR nonreciprocal behavior from Si photonic crystal designs

As a follow-up study, we employ the realistic dielectric function of Si and present the near-IR topological edge state characteristics of our proposed photonic crystal design. The interface design is identical to the one shown in the main text (see Figure 1(e)). In Figure S2(a), the resulting bulk energy band diagram of Si-based topological photonic design is shown. The band corresponding to the edge state is depicted with the solid red line and its electric field distribution along the photonic crystal at $k = 0$ depicts the confinement of the incident beam along the interface. It is important to note that, within the present design, we also incorporated the defects of both air hole (radius of 108 nm) and Si pillar defects (radius of 12 nm) at the interface which results in boosted self-induced nonreciprocity. Figures S2 (b) and (c) show the transmission response as the light is illuminated in backward and forward directions to the photonic crystal design, respectively. Higher nonreciprocal contrast is achieved with the increase of the incident beam power (see Figure S2(d)). The induced nonreciprocity contrast approaches a saturation value of ≈ 0.134 THz as the power of the incident beam is increased.

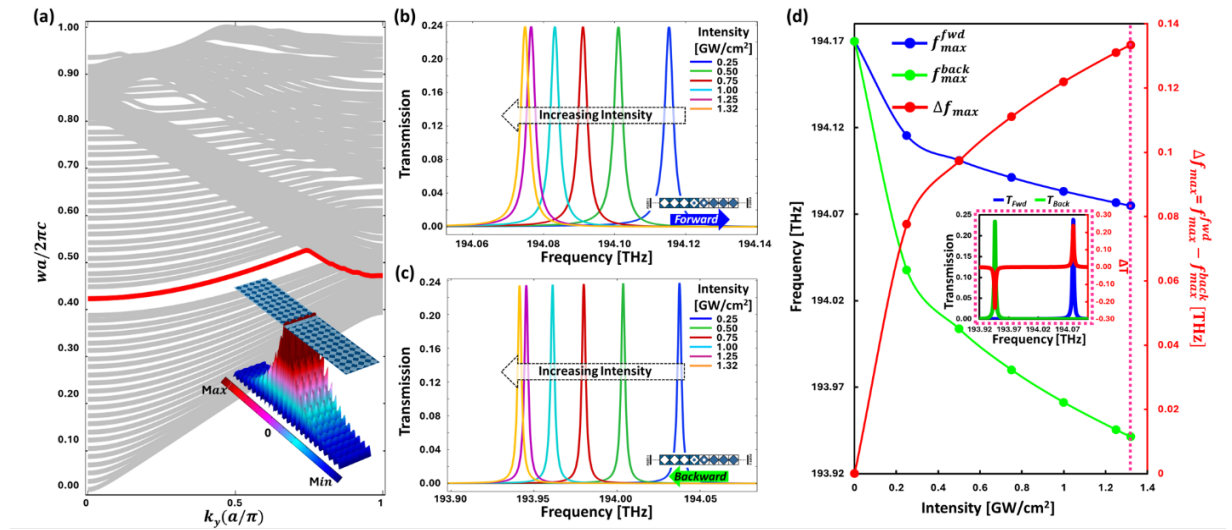

**Figure S2** (a) Dispersion band diagram of near-IR topological structure in the presence of both the interface and defects (air hole and Si pillar with radii of 108 nm and 12 nm, respectively). (b) The spectral evolution of transmission coefficient as a function of incident beam power under (b) forward and (c) backward illumination. (d) The trace of maximum values from both (b) $f_{max}^{back}$(green) and (c) $f_{max}^{fwd}$ (blue) are plotted as a function of incident beam intensity together with the tracing of maxima obtained from the difference between (b) and (c) ($\Delta f_{max} = f_{max}^{fwd} - f_{max}^{back}$, i.e., nonreciprocal frequency difference, red).

### 3. Nonreciprocal behavior from TiO$_2$ photonic crystal designs

Non-stoichiometry, defect, and imperfection free forms of metal oxides, ideally have minimal optical loss and extinction coefficient in the visible spectrum, making them transparent and insulating [2]-[4]. Thereby, this crucial link between optical response and the material properties in the metal-oxides make them suitable materials for the fabrication of photonic topological insulators. Furthermore, depending on their band gap and refractive index values, we can engineer a topological interface design that can operate in different parts of the spectrum.

In this section, we theoretically verify the operation of our photonic crystal design made of $TiO_2$ which can operate in two different spectral ranges: (a) near-IR and (b) UV. We utilized the spectroscopic ellipsometry-based measured complex dielectric function of amorphous state $TiO_2$ recently reported by Jolivet et al.[5]. To push the topological photonic gap into the near-IR spectral range, we select the lattice constant of our photonic crystal design to be 298 nm. In the case of $TiO_2$, the absence of loss within the near-IR spectral range yields strong coupling of the light at the interface and therefore a leakage resonance inside the photonic gap. When we look at the corresponding energy dispersion band diagram, we can easily identify the corresponding band for the edge state (see red solid line in Figure S3(a)). The spectral evolution of the transmission coefficient for backward and forward illumination cases is also plotted as a function of incident beam power in Figure S3(b). When the incident beam has high power, the spectral location of the resonance mode appears at larger wavelengths, and the amount of nonreciprocity of the topological edge mode becomes stronger. We also tested the performance of our photonic topological insulator design within the UV part of the optical spectrum always based on $TiO_2$. Note that the onset of absorption for $TiO_2$ starts at $\lambda \approx 376$ nm ($\approx 796$ THz). By adjusting the lattice constant of our periodic structure, we achieve efficient control of the edge state characteristics in the UV part of the spectrum with results shown in Figure S4.

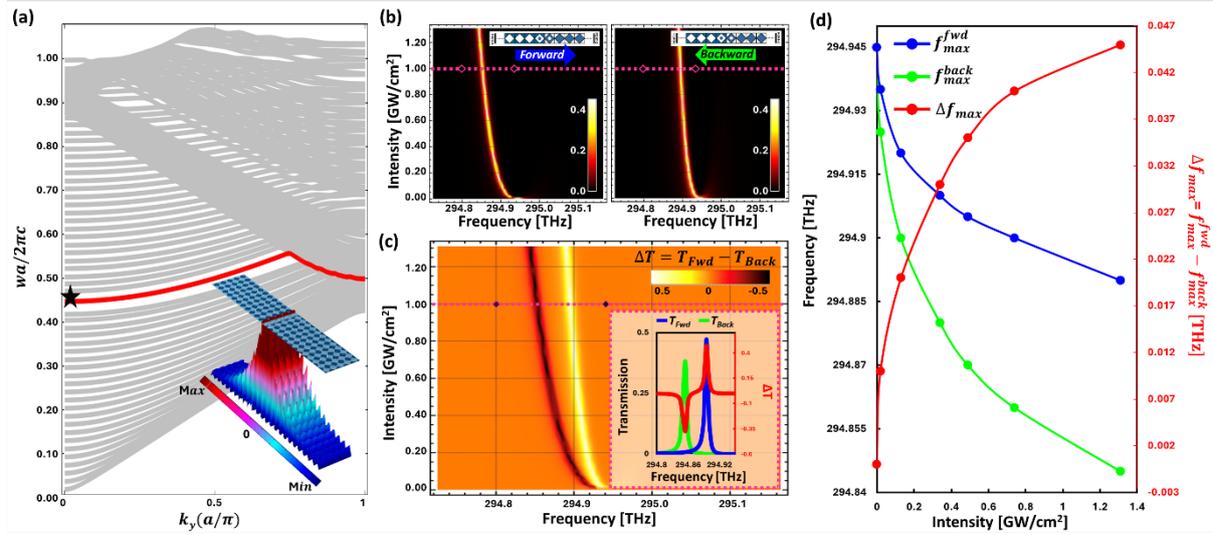

**Figure S3** (a) Dispersion band diagram of near-IR topological structure in the presence of both the interface and defects (air hole and TiO$_2$ pillar with radii of 98 nm and 11 nm, respectively). (b) The color density plot of the induced near-IR topological edge state spectra as a function of incident beam power under forward (left, $T_{Fwd}$) and backward (right, $T_{Back}$) illumination. (c) Nonreciprocal transmission contrast ($\Delta T = T_{Fwd} - T_{Back}$) as a function of incident beam power. The red dashed lines in both (b) and (c) are plotted as an inset in (c) which reveals strong, narrowband, and bisignate self-induced nonreciprocity (red solid line in the inset). (d) Traces of maxima resonance frequency difference derived by both (b) and (c) results and plotted as a function of incident beam intensity.

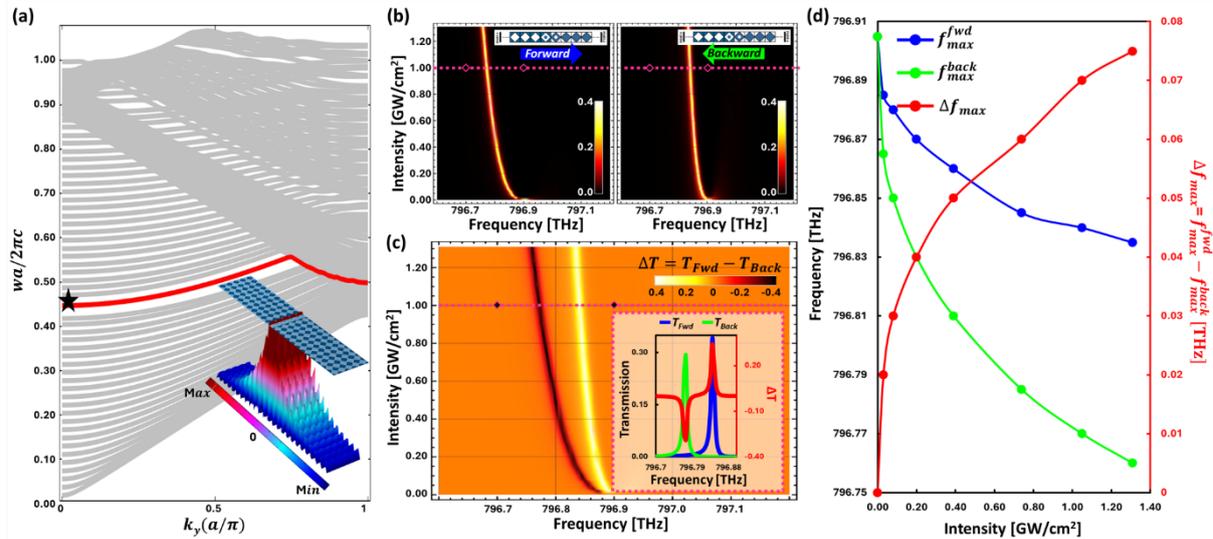

**Figure S4** (a) Dispersion band diagram of UV topological structure in the presence of both the interface and defects (air hole and TiO$_2$ pillar with radii of 32 nm and 4 nm, respectively). (b) The color density plot of the induced UV topological edge state spectra as a function of incident beam power under forward (left, $T_{Fwd}$) and backward (right, $T_{Back}$) illumination. (c) Nonreciprocal transmission contrast ($\Delta T = T_{Fwd} - T_{Back}$) as a function of incident beam power. The red dashed lines in both (b) and (c) are plotted as an inset in (c) which reveals strong, narrowband, and bisignate self-induced nonreciprocity (red solid line in the inset). (d) Traces of maxima resonance frequency difference derived by both (b) and (c) results and plotted as a function of incident beam intensity.

## 4. Electric field distribution induced at edge state

We theoretically study the location of induced edge state in the proposed superlattice structures and present their corresponding light coupling at the interface for two different cases: i) excitation at one edge perpendicular to the interface and ii) excitation at one edge parallel to the interface (see Figures S5 and S6, respectively). In both cases, we demonstrate the profile of the light coupled to the interface along the x-direction (see Figure S5(b) and Figure S6(b)) and y-direction (see Figure S5(c) and Figure S6(c)). Because the topological interface is along the x-direction, a sub-micrometer (approximately 4 adjacent unit cells) confinement in the strength of the electric field is observed along y-direction.

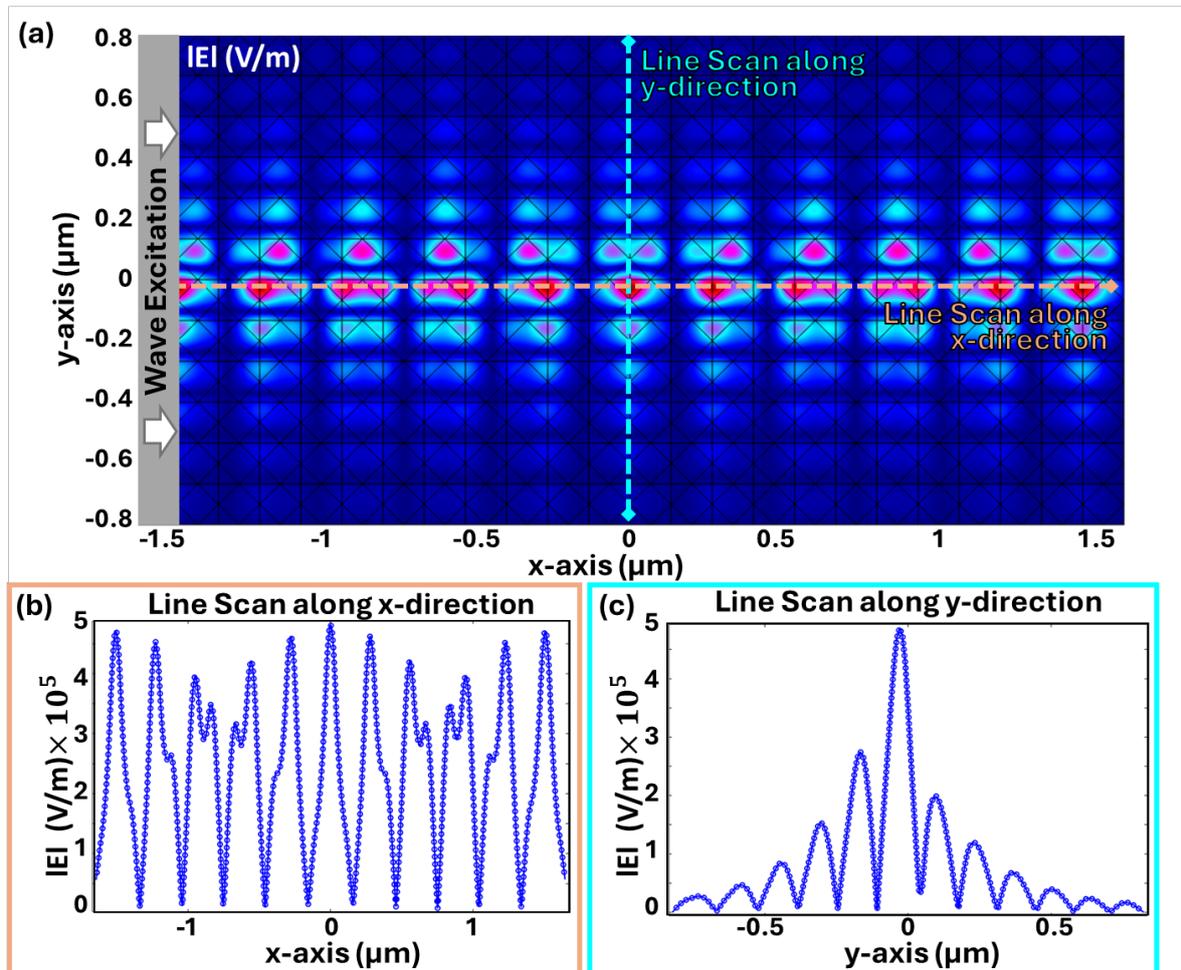

**Figure S5** (a) Coupling of electric field amplitude at a straight topological interface for wave excitation from left side of the structure. The confinement of the light coupled to the topological interface is demonstrated in figures (b) and (c) along the x- and y-direction, respectively. Because the interface is along the x-direction, the confinement in the strength of the electric field will be in the y-direction due to the edge state.

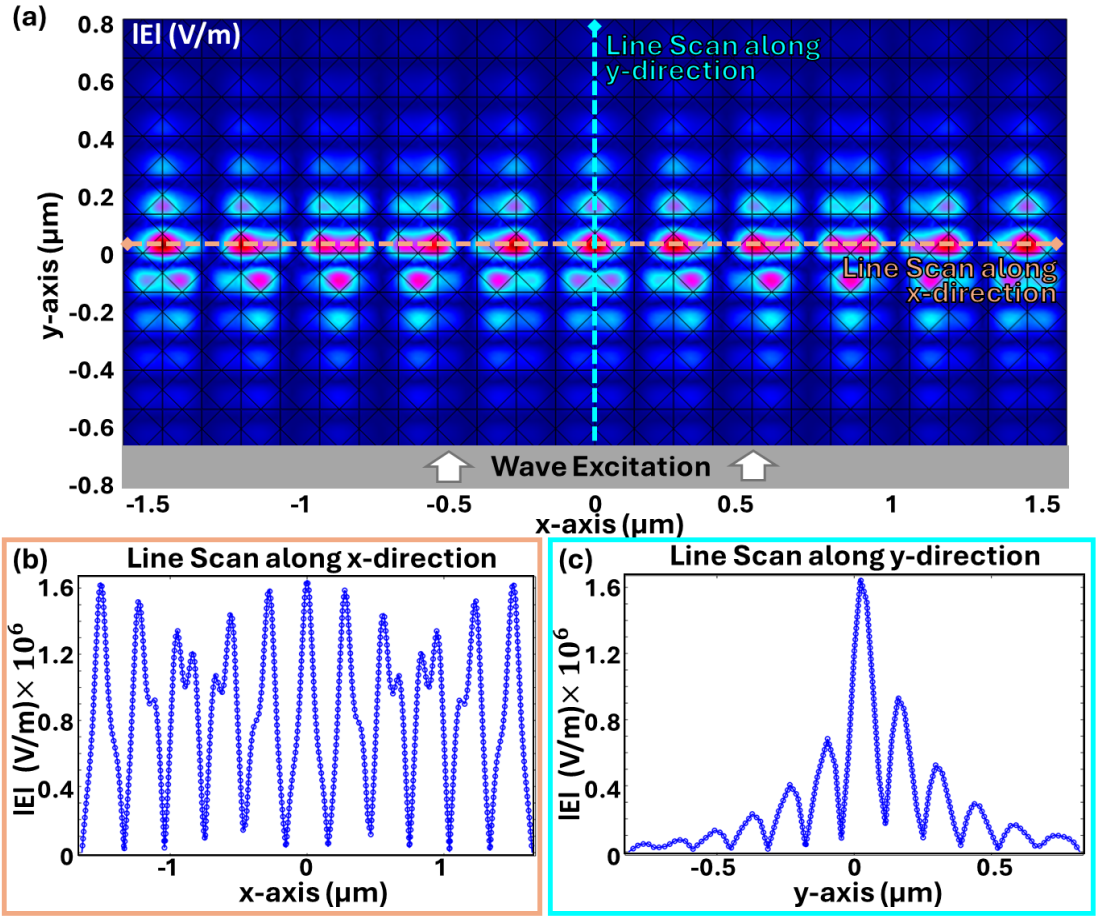

**Figure S6** (a) Coupling of electric field amplitude at a straight topological interface for wave excitation from bottom side of the structure. The confinement of the light coupled to the topological interface is demonstrated in figures (b) and (c) along the x- and y-direction, respectively. Because the interface is along the x-direction, the confinement in the strength of the electric field will be in the y-direction due to the edge state.